# An Aristotelian view on MR-based attenuation correction (ARISTOMRAC): combining the four elements

Matteo Cencini[1,2,3], Michela Tosetti[2,3] and Guido Buonincontri[2,3]

[1]*Dipartimento di Fisica, Università di Pisa, Pisa, Italy*, [2]*IRCCS Fondazione Stella Maris, Pisa, Italy*

[3]*Fondazione di Ricerca IMAGO7, Pisa, Italy*

**MR-based attenuation correction (MRAC) is important for accurate quantification of the uptake of PET tracers in combined PET/MR scanners. However, current techniques for MRAC usually require multiple acquisitions or complex post-processing to discriminate the different tissues. Inspired by the ancient Greeks, who believed that matter was made of the combination of four elements (earth, water, air and fire), we formulated a multi-component Magnetic Resonance (MR) Fingerprinting framework, where every voxel was considered a weighted combination of four base elements: bone, water, air and fat. We named our approach Aristotelian MR based attenuation correction (ARISTOMRAC). We used a 3D radial acquisition scheme at 1.5T, acquiring a transient-state spoiled acquisition with variable flip angles and echo times (TE), with the shortest TEs being ultra-short echo times (UTE). We simulated a multi-tissue MR signal model using the Bloch equations and used dictionary matching to extract tissue fraction maps for bone water and fat, while air fractions were obtained by thresholding the UTE parts of our acquisitions at higher spatial resolution. Compared to previous methods for MR-based Attenuation Correction (MRAC), our approach used a full multi-component signal model, including multiple tissues per voxel. For this reason, rather than reconstructing high resolutions images, MR data can be acquired more efficiently, directly at the resolution needed for PET attenuation maps. The ARISTOMRAC method allows to accurately estimate the air, water, bone and fat fractions (Concordance Correlation Coefficient = 0.81/0.91/0.98 for bone, water and fat respectively). Attenuation maps could be obtained in the head and neck with a single 1-minute acquisition.**

## I. Introduction

In many diagnostic applications[1], a correct photon attenuation correction is crucial for quantifying the uptake of a PET tracer. In PET/CT scanners photon attenuation is readily estimated from CT Hounsfield units, given the proportionality between Hounsfield units and electron density, which in turn gives the attenuation coefficient map. However, MRI image intensity is determined by proton density and relaxation times, which are not directly related to attenuation maps; therefore, in combined PET/MRI scanners, the attenuation correction relies on more sophisticated methods using information from MRI.

The two main classes of MR-based Attenuation Correction (MRAC) methods are atlas-based corrections and segmentation-based corrections [1]. In atlas-based corrections, MR images are co-registered to standard attenuation maps obtained from an atlas built upon several MRI/CT acquisitions. This approach has the advantage of requiring a single acquisition, but is prone to errors due to mis-registration and/or to inter-subject anatomical variability. On the other hand, the segmentation-based approach performs separation of the anatomical image in different regions, relying on the image intensity or on the location of the anatomical details; after this step, an attenuation coefficient is assigned to each tissue type. These techniques are more robust to anatomical variations with respect to atlas-based techniques but, since the tissue classification is usually based on T1-/T2- weighted images or Dixon images, many of them cannot distinguish between air and bone, due to the extremely low signal of ultra-short T2* in bone. Neglecting bone has detrimental effects on PET quantifications, leading to large underestimations around it [2]; to circumvent this problem, more sophisticated segmentation-based MRAC methods have taken advantage of acquisitions with ultra-short echo times (UTE), in order to separate the short T2* in bone from longer T2* in soft tissues. Such UTE acquisitions can also be used in combination with Dixon techniques, in order to extract bone, fat and water maps [3]. However, these approaches perform a hard segmentation of the bone tissues, thus requiring high resolution to obtain correct attenuation maps.

Here, we built upon the concept of Magnetic Resonance Fingerprinting (MRF) [4], [5] to develop a technique for brain MRAC. Briefly, an MRF experiment is based on a heavily undersampled transient-state acquisition, designed to obtain a unique signal evolution for each combination of tissue properties (i.e. Proton Density, T1 and T2). Then, these underlying tissue properties are recovered using an exhaustive search over a precompiled dictionary of possible signal evolutions obtained with a Bloch simulation. An interesting property of MRF is its intravoxel quantification capability. In fact, the transient-state signal evolution of a mixture of two tissues is completely different from the signal evolution of the two pure tissues; this capability can be exploited to perform sub-voxel quantification of tissue such as fat and water by using a two-component signal model [6]–[10]. Multicomponent estimations have also been previously described in the original MRF paper, which estimated segmentation of Grey Matter (GM), White Matter (WM) and Cerebrospinal Fluid (CSF) in the brain [4]. Other applications have demonstrated the discrimination between Intra/extracellular water and myelin water fractions [11], [12].

In this work, we aimed to obtain tissue segmentation within a single acquisition by including the bone in the multi-component model. Taking inspiration from the ancient Greeks, for whom all matter was a combination of four elementary substances (earth, water, air and fire), we modelled each voxel as a combination of water, fat, bone and air. Instead of trying to obtain a full set of parametric maps (i.e. T1, T2…) for each tissue type, we focused specifically on the tissue fractions. We assumed a limited set of parameters for each tissue class, thus mitigating the memory requirements to save the dictionary and decreasing the reconstruction time.

## II. METHODS

Our tissue fraction estimation builds on the concept of magnetic resonance fingerprinting (MRF), which compares transient-state signal evolutions to a pre-computed dictionary obtained simulating the Bloch equations [4]. In this work, we assumed that the MR signal evolution from a given voxel can be written as:

$$s(t) = (1-a) \cdot \rho \cdot m(t) = (1-a) \cdot \rho \cdot \sum_{i=1}^{3} w_i m_i(t) \quad (1)$$

where $s(t)$ is the total signal in a voxel, composed by a tissue fraction $(1-a)$ where $a$ is the air fraction, $\rho$ is a scaling factor dependent of receiver gain and local receiver sensitivity, and $m(t)$ is an $l2$-normalised evolution of transverse magnetization. We decomposed $m(t)$ in three different tissue components, where $w_i$ is the weight of the $i$-th component (with $w_i \in \mathbb{R}$, $\sum_{i=1}^{3} w_i = 1$) and $m_i(t)$ is the signal evolution for the $i$-th component.

### A. Dictionary Creation

A three-component MRF dictionary was created combining a water-only pool, a fat-only pool and a bone-only pool with weightings ranging from 0 to 1 (step size: 0.05), obtaining a dictionary of possible signal evolutions for each combination of the three tissue types. All the simulations were performed using the Extended Phase Graphs formalism [13]. This allowed to efficiently include the effect of gradient spoilers, RF pulse amplitudes and phases, phase evolutions due to off-resonance, and dephasing due to T2*. Off-resonance frequencies ranging from -300 to 300 Hz (step size = 5 Hz) were included in the model to allow fat-water discrimination in presence of field inhomogeneities. Singular Value Decomposition (SVD) of the signal evolution was applied retaining the first 10 singular values to reduce the dictionary storage burden and to improve matching speed [14]. Tissue fraction maps were obtained by inner product pattern matching between the acquired signals and the dictionary. Proton Density maps were obtained during the matching step as the scale factor between acquired and simulated signal evolution $l2$-norms [4]. The water pool was created assuming a coarse set of T1 relaxation times (500, 800, 900, 2500 ms, corresponding to white matter, glial/gray matter, muscle/skin and CSF; an additional entry with T1 = 700 ms was included to represent a lesion; values were taken from the *Brainweb* digital phantom[15]. A fixed T2* of 50ms was used for the water pool. The fat pool was created assuming a single T1/T2 value (230ms/70ms) and a multipeak spectrum (chemical shifts = 210, 159, -47, 236, 117, 23 Hz with relative amplitude = 0.62, 0.15, 0.1, 0.06, 0.03, 0.04 [16]). Finally, the bone pool was created assuming a bi-exponential model with a short T2 component (T1 = 100ms, T2 = 450μs) accounting for the 70% of bone signal and a long T2 component (T1 = 500ms, T2 = 4.1ms) accounting for the remaining 30% of the signal [17].

### B. Acquisitions

All the acquisitions were performed on a GE HDxt 1.5T scanner using an 8-ch receiver head coil. A 3D Radial center-out encoding scheme [18] with random permutation between spokes was used (total number of 3D spoke directions = 10530; field of view FOV = 25.6cm, matrix = 64x64x64), including both a gradient spoiler along the *z* direction in each repetition time (TR) and radiofrequency (RF) spoiling with quadratic phase increment of 117°.

Flip Angle (FA), Radiofrequency (RF) Phase and Echo Time/Repetition Time (TE/TR) were changed at each excitation with the aim of maximizing the differences between classes. The acquisition schedule is shown in Figure 1.

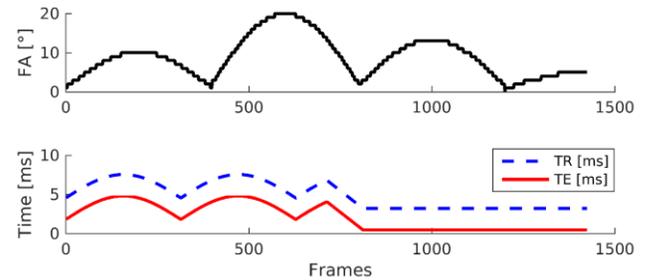

Fig. 1: Flip Angle and TE/TR acquisition schedule.

The first segment (frames from 1 to 811) had a variable TE with TEmin = 1.8ms to introduce off-resonance sensitivity (increasing Fat/Water discrimination) while minimising the bone signal [6]. The maximum TE used here was 4.8ms to reduce the T2* sensitivity. The second segment (frames from 812 to 1424) had a fixed TE = 0.455ms to achieve reliable signal from shorter T2 components in bone. Low flip angles (FAmax = 20°) were used to reduce potential confounding factors due to T1 differences within tissue classes, while RF spoiling was used to reduce confounding T2 effects due to the formation of stimulated echoes.

We compared an acquisition time of 105s to a shorter acquisition time of 49s, obtained truncating the acquisition.

*C. Air fraction estimations*

While it is feasible to separate water, fat and bone by looking at their different signal evolutions, air is more difficult to separate from tissue, since air has no MR signal. Here, we used a threshold on data reconstructed at a higher nominal spatial resolution, to determine whether any location has signal or not, then reduced the spatial resolution to obtain air fractions.

We summed the undersampled MRF k-space volumes acquired with UTE (frames 812 to 1424) across the time dimension, reconstructing one image at twice the nominal spatial resolution (2mm iso). Such image has signal from both tissues and bone and can be used to segment air. Here, we simultaneously bias corrected along the z direction and obtained a binary mask by applying a threshold in each axial partition of 3.5 times the median of the signal in each partition. At this point, images were smoothed to a 4mm resolution, creating a range of air fractions taking into account partial volumes.

*D. Phantom experiment*

To validate the technique, a phantom, consisting of a bovine bone surrounded by a 0.6% agar gel was scanned. Four vials, filled with emulsions of water and vegetable oil (nominal volume fractions: 28%, 56%, 80%, 100%) were also included to assess the fat fraction quantification capability of the technique. To obtain ground truth values for the tissue fraction, a Cartesian T1-weighted 3D image was acquired (3D SPGR TR = 14.3 ms, TE = 4.4 ms, Flip Angle = 10°, matrix = 256x256x146, in-plane FOV = 25.6x25.6x19.0 cm corresponding to an in-plane resolution of 1mm²). This image was then segmented by intensity thresholding and manually classified according to the nominal tissue fractions. The maps were downscaled by Gaussian smoothing followed by a nearest neighbour interpolation to match the ARISTOMRAC resolution. Finally, the ground-truth maps were co-registered to the ARISTOMRAC maps using FSL [19]. Average tissue fraction for both ground truth and ARISTOMRAC were acquired within each region of the phantom (shown in Figure 2) and used to calculate the Concordance Correlation Coefficient (CCC) [20] between measured and nominal values on the central 3 slices of our acquisition, covering the phantom:

$$\rho_c = \frac{2S_{12}}{S_1^2 + S_2^2 + (\bar{Y}_2 - \bar{Y}_1)^2} \quad (2)$$

where $\bar{Y}_j$, $S_j^2$ and $S_{12}$ are $\bar{Y}_j = \frac{1}{n}\sum_{i=1}^{n} Y_{ij}$, $S_j^2 = \frac{1}{n}\sum_{i=1}^{n}(Y_{ij} - \bar{Y}_j)^2$ and $S_{12} = \frac{1}{n}\sum_{i=1}^{n}(Y_{i1} - \bar{Y}_1)(Y_{i2} - \bar{Y}_2)$, $Y_1, Y_2$ are the tissue fraction values for ground truth (j=1) and ARISTOMRAC (j=2) and $n$ is the number of regions.

To evaluate the ability of the technique to discriminate air from tissue, the ground truth air map was converted in a binary mask (including each voxel with air fraction > 0.01) and the Dice similarity index $D$ [21], [22] was calculated between the ARISTOMRAC air mask and this ground truth air mask:

$$D = \frac{2TP}{(2TP + FP + FN)} \quad (3)$$

where *TP*, *FP* and *FN* are the numbers of True Positive, False Positive and False Negative.

We repeated this analysis for both the full acquisition (105s) and the truncated one (49s).

To assess the capability of the technique to estimate tissue partial volumes, we downscaled both ground truth and ARISTOMRAC maps to 8 and 16 mm isotropic resolutions. Then, we calculated the differences between ARISTOMRAC and ground truth values and we computed mean and standard deviation of these differences. This analysis was restricted to voxels with a tissue fraction higher than 0.5. In addition, the Dice similarity between air maps and ground truth were calculated for each resolution. The results were reported as histograms for each tissue type and each resolution.

To evaluate the accuracy of the attenuation coefficients, the attenuation map was calculated as a weighted sum of the attenuation coefficient of water, fat and bone (respectively 0.09, 0.1 and 0.172 cm$^{-1}$ [3]), where the weighting coefficients were given from tissue fractions. Mean and standard deviations of attenuation coefficients were compared for each ROI between the ARISTOMRAC and ground truth attenuation maps. In addition, attenuation profiles were acquired along lines crossing each vial, to evaluate accuracy of attenuation coefficients at air/tissue interfaces.

*E. In vivo experiment*

To test the *in vivo* capability of the technique, a healthy human volunteer was acquired in accordance with the protocol for volunteer MR experiments. The ARISTOMRAC attenuation map was derived as in the phantom experiment. Following the results of the phantom experiments, in vivo we only investigated a short acquisition of 49s. To provide a comparison for this attenuation map, a T1-weighted anatomic image was acquired (3D SPGR TR = 14.3 ms, TE = 4.4 ms, Flip Angle = 10°, matrix = 256x256x256, 1mm³ isotropic resolution). This image was segmented using the MARS (Morphologically and Anatomically accuRate Segmentation) extension for version 8 of Statistical Parametric Mapping (SPM8) [23], [24] and the resulting tissue fractions were used to calculate a reference attenuation map. This map was finally downscaled to match ARISTOMRAC resolution and co-registered to ARISTOMRAC attenuation map [25]. Mean and standard deviation of attenuation coefficients were compared between ARISTOMRAC and reference acquisitions in three ROIs. The ROIs were determined by the SPM segmentation of the reference acquisition. ROIs used for this analysis were brain, scalp and skull. Dice similarity between ARISTOMRAC and reference was calculated as in the phantom experiment.

III. RESULTS

Dictionary creation took 56s on an Intel® Xeon® processor E5-2600 v4. After SVD compression, dictionary size was 10MB. Notice that this step must performed once for a given acquisition schedule. Image reconstruction and pattern matching were performed in 97s.

Using ARISTOMRAC, we successfully obtained tissue fraction maps in vitro, as shown in Figure 2.

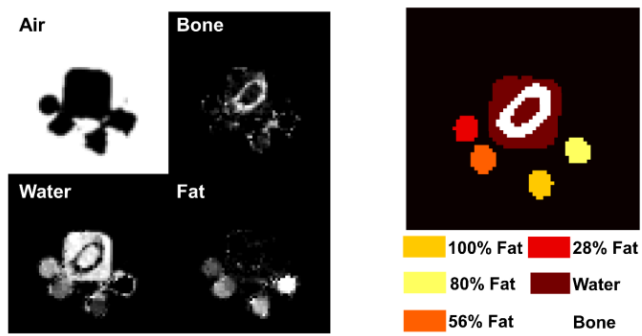

Fig. 2: Tissue fraction maps for the phantom obtained with a 49 s acquisition (left) and nominal values (right). Tissue ROIs were defined as regions with a ground truth tissue fraction > 0.5. Bone ROI was defined as the region with a ground truth bone fraction > 0.5.

The bone was correctly classified as a tissue while the background was classified as air, and the resulting Dice similarity index was 0.96/0.97 (for the 105s and the 49s acquisitions respectively). We observed good agreement between the measured tissue fractions and the nominal values: CCC was 0.87/0.81 (for the 105 s and the 49s acquisitions respectively) for bone fraction, 0.96/0.91 for water fraction and 0.99/0.98 for fat fraction. The major source of inconsistency between nominal and measured values was given by the mixed tissue-air voxels.

Histograms in Figure 3 demonstrate the capability of our technique to extract partial volume fractions within the single voxel.

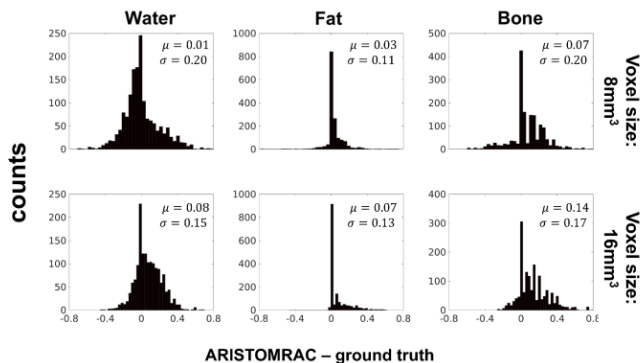

Fig. 3: Histograms comparing the estimated tissue fractions with the ground truth for artificial resolution of 8mm isotropic (top) and 16mm isotropic (bottom). Mean and standard deviation of the errors are reported for each histogram.

For an artificial resolution of 8mm isotropic, quantification remained accurate: the mean error was less than 0.07 (which was close to the step size of the tissue fractions), while the standard deviation of the error was less than 0.20. By further degrading the resolution to 16mm isotropic, the accuracy of the measurement decreases, the mean error being 0.07 for fat, 0.08 water and 0.14 for bone. As shown by the asymmetry of the histograms, ARISTOMRAC tends to slightly overestimate tissue fraction values. This phenomenon become more evident for extremely low resolution, and should be interpreted as classification of mixed air-tissue voxels as pure tissue voxels. This is supported by the behaviour of the Dice similarity index between ground truth air and estimated air masks: in fact, Dice index was 0.94 for the 8mm$^3$ resolution and 0.90 for the 16mm$^3$ resolution. In Table 1 mean and standard deviation of attenuation coefficients for each ROI of the phantom are reported for both ARISTOMRAC and ground truth. Attenuation profiles across the vials are shown in Figure 4.

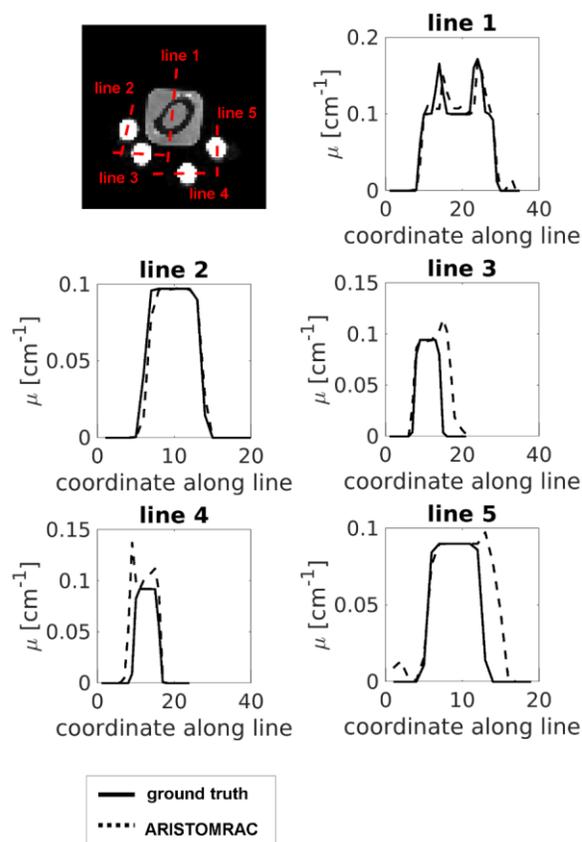

Fig. 4: comparison between ground truth and ARISTOMRAC attenuation profiles across each vial of the phantom.

It can be seen that ARISTOMRAC tends to overestimate tissue fractions in some cases, as near the fat vials where the chemical shift artefacts induce displacement of the signal.

| ROIs | μ ground truth [cm$^{-1}$] | μ ARISTOMRAC [cm$^{-1}$] |
|---|---|---|
| Water | 0.10±0.01 | 0.10±0.02 |
| 28% fat | 0.09±0.01 | 0.09±0.02 |
| 56% fat | 0.09±0.01 | 0.09±0.02 |
| 80% fat | 0.08±0.01 | 0.10±0.01 |
| 100% fat | 0.08±0.01 | 0.09±0.02 |
| Bone | 0.14±0.02 | 0.13±0.02 |

Table 1 comparison between ground truth and ARISTOMAC attenuation values.

In the *in vivo* experiment (shown in Figure 5), mouth and nasal sinus were correctly classified as air; the skull is clearly visible in the bone fraction map, subcutaneous fat is correctly reconstructed in the fat fraction map and the rest of the tissues are mostly classified as water, as expected.

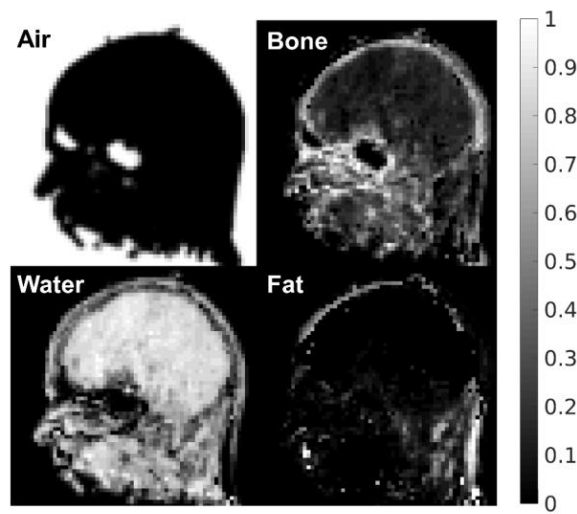

Fig. 5: Tissue fractions in vivo in a human head obtained with ARISTOMRAC technique (acquisition time: 49s).

This is reflected in the good visual agreement between the ARISTOMRAC attenuation map and the one obtained by the segmentation of the T1-weighted image (Figure 6).

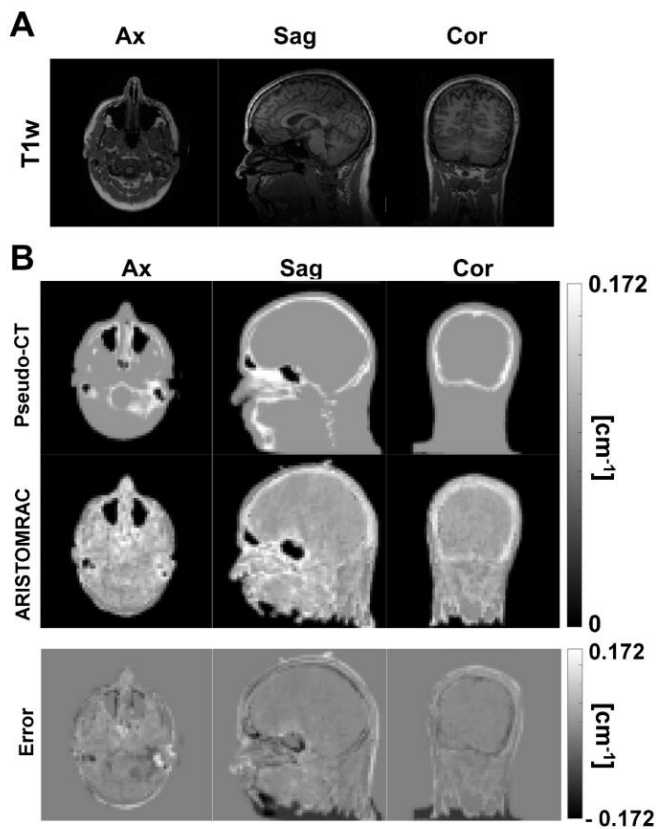

Fig. 6: A) T1-weighted anatomical image used to obtain reference attenuation map. B) Top row: attenuation map obtained from the ARISTOMRAC tissue fractions and from the segmentation of the T1-weighted image (first column: axial view; second column: sagittal view; third column: coronal view). Bottom row: difference between ARISTOMRAC and reference attenuation maps.

Dice similarity between ARISTOMRAC and reference acquisition in vivo was 0.98. Mean and standard deviation of attenuation values in the brain, scalp and skull are reported in Table 2 for both ARISTOMRAC and reference acquisitions.

| ROI | µ reference [cm$^{-1}$] | µ ARISTOMRAC [cm$^{-1}$] |
|---|---|---|
| Brain | 0.100± 0.001 | 0.110± 0.010 |
| Scalp | 0.090± 0.025 | 0.094± 0.042 |
| Skull | 0.130± 0.024 | 0.120± 0.033 |

Table 2 comparison between reference and ARISTOMRAC attenuation values.

In addition to the attenuation maps, we also report the 3D B0 map and a 3D volumetric image in Figure 7. Although these are not useful for MRAC, they are obtained as byproducts of our analysis and may be useful within a full MR protocol.

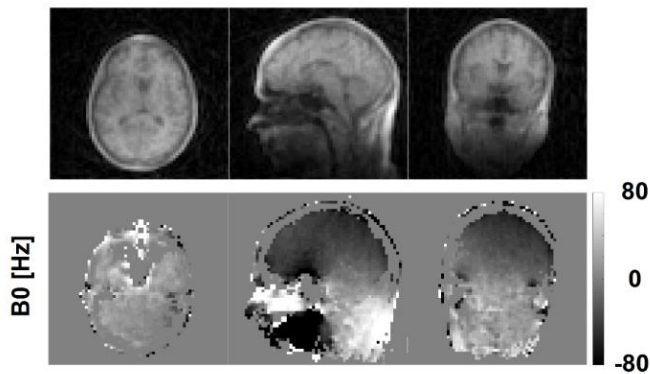

Fig. 7: The 3D localiser images (top row) and 3D B0 map (bottom row) obtained with ARISTOMRAC.

## IV. Discussion and Conclusions

In the present work, we demonstrated a novel technique to perform MR attenuation correction for PET photon quantification. We were able to obtain tissue fraction maps for air, bone, fat and water of an entire volume in a short scanning time (49s) *in vivo* and we were able to obtain a map of attenuation coefficient of a human head and neck. Importantly, the multicomponent quantification capability of our technique allowed to acquire the image directly at the PET resolution. This property allowed to save acquisition time with respect to other multi-echo steady-state acquisitions in which the bone structures are identified with hard segmentation that requires high resolution acquisitions [3].

In comparison, MRAC techniques from Keereman et al [2]. required a 3.5 min acquisition, Berker et al. required 214 s and Ladefoged et al. required approximately 100 s [26].

Our work leveraged the intrinsic multi-component capabilities of MR Fingerprinting. A similar segmentation approach has been previously exploited in the original MRF paper, in which a least-square minimization was used to perform sub-voxel decomposition of Grey Matter (GM), White Matter (WM) and Cerebrospinal Fluid (CSF) in the brain [4]. Instead of representing the segmentation as a least-square problem, here we use an exhaustive search over discretized combination of the pure tissue signals. In this way, tissue fractions are automatically constrained to be real and their sum is normalized to 1 [27]. Another work relying on dictionary-based sub-voxel quantification was the one by Hamilton et al. [11], in which T1, T2, signal fraction and exchange rate were determined for two components (Intracellular and Extracellular Water), resulting in a very large dictionary and long reconstruction time. In our technique we used a much simpler approach, focusing specifically on the tissue fractions and adopting a minimal description of the tissue classes, thus mitigating the memory requirements to save the dictionary and decreasing the reconstruction time.

With our method, we obtained accurate attenuation maps despite some errors in tissue fraction maps, like some residual bone fraction in soft tissues, leading to overestimation of the attenuation values. However, we did not specifically optimize the tissue T1, T2 and chemical shift combinations in our model. Rather than tuning these values, we used values taken from well-established literature studies. As the resulting tissue fractions depend on the choice of these parameters, fine tuning may permit to obtain even more accurate and robust results. For instance, we found that decreasing the T2* of water pool from 50ms to 10ms leads to a significant decrease in the associated biases in the μ map, due to a more correct estimation of bone. This is reported in Figure 8, where it can be seen that the residual bone signal in soft tissues is nulled with the tuned parameters.

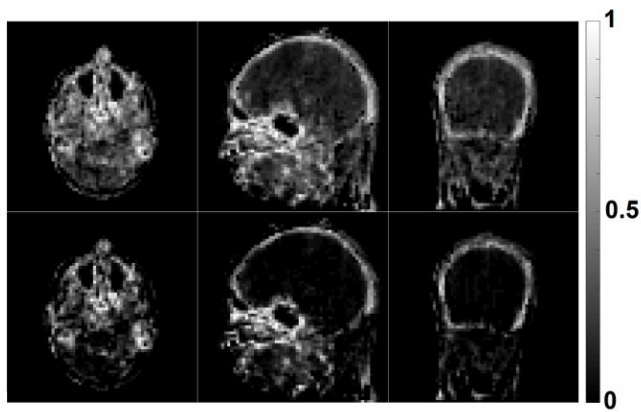

Fig. 8: Comparison between ARISTOMRAC bone fractions with water T2* = 50ms (top) and 10ms (bottom) in one subject obtained in a post-hoc analysis, showing that fine-tuning of the model parameters may result in better estimation.

Further studies acquiring data in a cohort of subjects, may permit to fine-tune the model parameters to achieve the highest accuracy and precision for MRAC in a specific application. Also, it would be possible to optimise the acquisition, using established methods for maximising encoding capabilities of transient-state acquisitions [28], [29]. Low T2* values of bone are the main discriminant in comparison with soft tissue, as they induce signal loss by dephasing for the longer TEs, while both T1 and T2* contribute to the *l2* norm of the signal evolutions, hence impacting fraction estimations.

While our tissue segmentation model used a novel acquisition and a three-component MR Fingerprinting model, air segmentation was based on a high-resolution UTE image obtained with a part of our acquisition frames, then downsampled to match the other lower-resolution tissues and obtaining air fractions. As we used a high-resolution UTE image, it would be possible to use ARISTOMRAC in combination with any established air segmentation techniques in literature such as [30]. In addition to the tissue fractions for the attenuation maps, our method also produced a 3D anatomical image and a 3D B0 map of soft tissues. These can be used for localization and/or high order shimming procedures.

The major advantage of the ARISTOMRAC technique is that it incorporates some advantages of both segmentation and atlas based MRAC techniques. Like segmentation MRAC, ARISTOMRAC can directly extract the tissue fractions from the data, hence it does not suffer from errors due to mis-registration with the atlases; further, like atlas-based techniques, it only requires a single acquisition to obtain the attenuation maps.

Despite several advantages, the technique has some limitations. The phantom experiment showed that the bone signal may be masked from the other two components for low bone fraction values. Moreover, this technique was designed specifically for the head and neck. Application in other body areas may require a more thorough study of inter-patient variability as well as methods to deal with physiological motion[31], [32].

Another limitation of this study is the lack of a comparison with gold standard CT-based Attenuation Correction. However, the measured tissue fractions in the phantom experiment reflect the nominal composition of the phantom, giving a first demonstration of ARISTOMRAC attenuation maps.
Future extensions of this work are required to perform such validation and to test the impact of the correction on tracer quantification in PET images, as well as to optimize the trajectory and acquisition schedule.

In conclusion, the ARISTOMRAC technique represents a promising approach to perform MRAC as an alternative to existing techniques.